\documentclass[aps, prl, twocolumn, groupedaddress, 10pt, showpacs ,superscriptaddress]{revtex4-1}

\usepackage{graphicx}
\usepackage{color}
\usepackage{amsfonts,amsmath,amsthm,amssymb,amscd}
\usepackage{dsfont}
\usepackage{mathrsfs}
\usepackage{mathtools}
\usepackage{yfonts}

\def\epp{\: .}
\def\epc{\: ,}
\def\lam{\lambda}
\def\blam{{\boldsymbol{\lambda}}}
\def\brho{{\boldsymbol{\rho}}}
\def\limth{\lim\nolimits_\text{th}}
\def\cO{\mathcal{O}}

\def\bra#1{\mathinner{\langle{#1}|}}
\def\ket#1{\mathinner{|{#1}\rangle}}
\newcommand{\RaR}{\mathbb{R}}
\newcommand{\CC}{\mathbb{C}}
\def\ua{\uparrow}
\def\da{\downarrow}

\def\one{\mathds{1}}
\def\squote#1{\lq{#1}\rq}
\newcommand{\be}{\begin{equation}}
\newcommand{\ee}{\end{equation}}
\newcommand{\bea}{\begin{eqnarray}}
\newcommand{\eea}{\end{eqnarray}}
\def\nn{\nonumber\\}
\def\fr#1{(\ref{#1})}

\begin{document}

\title{Complete Generalized Gibbs Ensemble in an Interacting Theory}

\author{E. Ilievski}
\affiliation{Institute for Theoretical Physics, University of Amsterdam, Science Park 904, 1098 XH Amsterdam, The Netherlands}

\author{J. De Nardis}
\affiliation{Institute for Theoretical Physics, University of Amsterdam, Science Park 904, 1098 XH Amsterdam, The Netherlands}

\author{B. Wouters}
\affiliation{Institute for Theoretical Physics, University of Amsterdam, Science Park 904, 1098 XH Amsterdam, The Netherlands}

\author{J.-S. Caux}
\affiliation{Institute for Theoretical Physics, University of Amsterdam, Science Park 904, 1098 XH Amsterdam, The Netherlands}

\author{F. H. L. Essler}
\affiliation{The Rudolf Peierls Centre for Theoretical Physics, University of Oxford, Oxford, OX1 3NP, United Kingdom}

\author{T. Prosen}
\affiliation{Faculty of Mathematics and Physics, University of Ljubljana, Jadranska 19, SI-1000 Ljubljana, Slovenia}

\date{\today}

\begin{abstract}
In integrable many-particle systems, it is widely believed that the
stationary state reached at late times after a quantum quench can be
described by a generalized Gibbs ensemble (GGE) constructed from their
extensive number of conserved charges. A crucial issue is then to
identify a complete set of these charges, enabling the GGE to provide
exact steady state predictions.  
Here we solve this long-standing problem for the case of the spin-1/2
Heisenberg chain by explicitly constructing a GGE which uniquely fixes
the macrostate describing the stationary behaviour after a general
quantum quench. A crucial ingredient in our method, which readily
generalizes to other integrable models, are recently discovered
quasi-local charges. As a test, we reproduce the exact
post-quench steady state of the N{\'e}el quench problem obtained previously by means
of the Quench Action method. 
\end{abstract}


\maketitle

\paragraph*{Introduction.}
Understanding and describing the equilibration of isolated many-particle
systems is one of the main current challenges of quantum physics. The
presence of higher conserved charges (above the Hamiltonian) is linked
to the absence of full relaxation to a 
thermalized state; the conjectured appropriate framework to
characterize the steady state properties in such a situation is the
Generalized Gibbs Ensemble (GGE) \cite{2007_Rigol_PRL_98}, in which
all available charges are ascribed an individual `chemical potential'
set by the initial conditions, and the steady state is the maximal
entropy state fulfilling all the constraints associated to the conserved charges
\cite{2006_Rigol_PRA_74,2006_Cazalilla_PRL_97,2007_Calabrese_JSTAT_P06008,2008_Cramer_PRL_100,2008_Barthel_PRL_100,2010_Fioretto_NJP_12,2011_Calabrese_PRL_106,2012_Calabrese_JSTAT_P07016,2013_Fagotti_PRB_87,2012_Caux_PRL_109,2012_Essler_PRL_109,2011_Cassidy_PRL_106,2013_Caux_PRL_110,2013_Collura_PRL_110,2013_Mussardo_PRL_111,2013_Pozsgay_JSTAT_P07003,2013_Fagotti_JSTAT_P07012,2013_Kormos_PRB_88,2014_Fagotti_PRB_89,2014_Wouters_PRL_113,2014_Pozsgay_PRL_113,2014_Kormos_PRA_89,2014_DeNardis_PRA_89,2014_Sotiriadis_JSTAT_P07024,2015_Essler_PRA_91,2015_Mestyan_JSTAT_P04001}. 
The basic idea underlying the GGE is as follows. Let $H\simeq H^{(1)}$ be the
Hamiltonian of an integrable model, and $\{H^{(n)}\}$ a set of conserved
charges fulfilling $[H^{(n)},H^{(m)}]=0$. The situation we are interested in
is that of a quantum quench, where we initially prepare our system in
the ground state $|\Psi(0)\rangle$ of a local Hamiltonian
$H_0$ and then consider unitary time evolution with respect to our
integrable Hamiltonian 
\be
|\Psi(t)\rangle=e^{-iHt}|\Psi(0)\rangle.
\ee
We assume that we are dealing with a generic case, where in the
thermodynamic limit $|\Psi(0)\rangle$ cannot be expressed as a linear
combination of any finite number of eigenstates of $H$. At late times
after the quench expectation values of local operators approach
stationary values 
\be
\langle {\cal O} \rangle_\Psi=\lim_{t\to\infty}\langle\Psi(t)|{\cal O}|\Psi(t)\rangle.
\ee
The GGE hypothesis asserts that these expectation values can be
calculated as $\langle {\cal O} \rangle_\Psi={\rm Tr}(\hat{\varrho}_{\rm GGE}{\cal O})$ from a
statistical ensemble with a density matrix
\be
\hat{\varrho}_{\rm GGE}=\frac{1}{Z}\exp\left[-\sum_{n}\beta_n H^{(n)}\right].
\label{rhoGGE}
\ee
Here $Z$ is a normalization, and the Lagrange multipliers $\beta_n$
are fixed by the initial conditions 
\be
\limth\frac{{\rm Tr}(\hat{\varrho}_{\rm GGE}H^{(n)})}{N}
=\limth
\frac{\langle\Psi(0)|H^{(n)}|\Psi(0)\rangle}{N},
\label{GGE_fixbeta}
\ee
where $N$ is the system size and $\limth$ denotes the thermodynamic limit $N \rightarrow \infty$. Eqns.~\fr{GGE_fixbeta} are a direct
consequence of the fact that $H^{(n)}$ are conserved charges.
While the GGE hypothesis has been successfully verified for
many systems mappable to free particles, in interacting theories such
as the spin-1/2 Heisenberg XXZ chain the question arises,
\emph{precisely which charges need to be included in \fr{rhoGGE}}. In
Refs. \cite{2013_Pozsgay_JSTAT_P07003,2013_Fagotti_JSTAT_P07012,2014_Fagotti_PRB_89} a GGE based on the known 
conserved local charges \cite{KorepinBOOK} was constructed
and used to determine steady-state averages of observables
\cite{2014_Fagotti_PRB_89}. Subsequent analyses 
\cite{2014_Wouters_PRL_113,2014_Pozsgay_PRL_113} by the Quench Action
(QA) approach \cite{2013_Caux_PRL_110} demonstrated that this
GGE fails to predict the correct steady state properties. This failure
was shown to be related to the existence of bound states
\cite{2014_Wouters_PRL_113} (see also 
\cite{2014_Pozsgay_JSTAT_P09026,Goldstein_arxiv14054224}), which are known to be a generic
feature in quantum integrable models. These results posed the question
whether the GGE is conceptually faulty, or whether there could exist
hitherto unknown charges that need to be taken into account in its
construction. 

In this Letter, we settle this issue by explicitly showing how to
repair the GGE in Heisenberg chains, by complementing it with
recently-discovered additional families of conserved charges
\cite{Ilievski_arxiv150605049}. Crucially, these ``quasi-local''
charges fulfil a weaker form of locality than the
previously known ones. We derive a set of fundamental identities
between the initial-state expectation values of these charges, and the
density functions characterizing the steady state. An explicit test of
our construction is provided by a quantum quench from the N{\'e}el
state to the $XXZ$ chain: we demonstrate that our GGE correctly
recovers the stationary state, the form of which is known exactly from the
QA approach \cite{2014_Wouters_PRL_113,2014_Brockmann_JSTAT_P12009}.
In this way we completely resolve the above-mentioned conundrum. 
Our construction shows that quasi-local conserved charges are in fact
crucial for understanding the non-equilibrium dynamics of quantum
integrable models. 

\paragraph*{Anisotropic spin-1/2 Heisenberg chain.} 
We shall consider a completely generic quench protocol from an initial
pure wavefunction $\ket{\Psi_0}$, which is unitarily time evolved
according to the Hamiltonian
\begin{equation}\label{eq:Hamiltonian_XXZ}
	H = \frac{J}{4}\sum_{j=1}^{N}\left[\sigma_{j}^{x}\sigma_{j+1}^{x}+\sigma_{j}^{y}\sigma_{j+1}^{y}+
	\Delta ( \sigma_{j}^{z}\sigma_{j+1}^{z}-1)\right].
\end{equation}
Here $J>0$, $\sigma^\alpha_j$, $\alpha = x, y, z$ are Pauli matrices
acting on spin-$1/2$ degrees of freedom, and we consider anisotropy
values in the regime $\Delta=\cosh{(\eta)} \geq 1$.  
The Hamiltonian (\ref{eq:Hamiltonian_XXZ}) can be diagonalized by
Bethe Ansatz \cite{1931_Bethe_ZP_71,1958_Orbach_PR_112}. Imposing
periodic boundary conditions, energy eigenstates $|\blam\rangle$ with
magnetization $S^z_{\text{tot}} = \frac{N}{2} - M$ are labeled by a
set of rapidities $\blam = \{ \lambda_k \}_{k=1}^M$ satisfying the
Bethe equations  
$\left(\frac{\sin(\lambda_j+i\eta/2)}{\sin(\lambda_j-i\eta/2)}\right)^N=
-\prod_{k=1}^M\frac{\sin(\lambda_j-\lambda_k+i\eta)}{\sin(\lambda_j-\lambda_k-i\eta)}$,
$j=1,\ldots,M$. The momentum and energy of a Bethe state are
$P_\blam =\sum_{j=1}^M p(\lam_j)$, 
$\omega_{\blam} = \sum_{j=1}^M e(\lam_j)$
where $p(\lam) = i \ln\left[\frac{\sin(\lambda - i \eta/2)}{\sin(\lambda + i \eta/2)}\right]$ and
$e(\lam) = - J \pi  \sinh(\eta) a_1(\lambda)$, where
\begin{equation}
\label{eq:an}
a_n(\lambda) = \frac{1}{2\pi}\frac{2\sinh(n\eta)}{\cosh(n\eta) -\cos{(2\lambda)}}.
\end{equation}
Solutions $\blam$ to the Bethe equations are closed under complex
conjugation and consist of so-called strings
$\lam^{n,a}_\alpha=\lam^n_\alpha+\tfrac{i\eta}{2}(n+1-2a) +
i\delta^{n,a}_\alpha$, $a=1,\ldots,n$ and $\lam^n_\alpha \in \RaR$. Here index $\alpha$ enumerates a string, $n$ is the string length,
$a$ counts rapidities inside a given string and deviations $\delta^{n,a}_\alpha$ are (for the majority of states)
exponentially small in system size
\cite{GaudinBOOK,KorepinBOOK,TakahashiBOOK}. The string centers
$\lam^n_\alpha$ lie in the interval $[-\pi/2,\pi/2)$. In the
thermodynamic limit $N \to \infty$ with $M/N$ fixed one can describe a state not in terms of individual
rapidities, but rather in terms of a set of functions
$\boldsymbol{\rho} = \{ \rho_n \}_{n=1}^\infty$ representing string
densities (see Supplementary Material (SM) for more info). 

\paragraph{Ultra-local GGE treatment.}
Exactly-solvable Hamiltonians such as (\ref{eq:Hamiltonian_XXZ}) can
be embedded  \cite{KorepinBOOK} in a commuting family $[T(\lambda),  T(\lambda')] = 0$ of
transfer matrices (defined in (\ref{Tsz})). The Hamiltonian and an infinite
number of mutually commuting conserved charges are obtained by
\be
H^{(n)} = \frac{i}{n!}\partial^{n}_{\lambda}
 \ln T(-i\lambda)\Big|_{\lambda=\frac{i \eta}{2}}
\label{ultralocal}
\ee
with the Hamiltonian reading $H = \frac{J\sinh(\eta)}{2} H^{(1)}$.
These charges are ultra-local in the sense that they can be written as
$H^{(m)} = \sum_{j=1}^N h_j^{(m)}$, where the operators $h_j^{(m)}$ act
nontrivially on a block of at most $m$ sites adjacent to $j$. The GGE
constructed in \cite{2013_Pozsgay_JSTAT_P07003,2013_Fagotti_JSTAT_P07012} was of the form
\fr{rhoGGE}, \fr{GGE_fixbeta} with charges \fr{ultralocal}. The
initial values $h^{(n)}=\lim_{\rm th}N^{-1}\langle\Psi(0)|H^{(n)}|\Psi(0)\rangle$
of the conserved charges are conveniently encoded
in the generating function \cite{2013_Fagotti_JSTAT_P07012}
\bea
\label{eq:defgeneratingfunction}
\Omega^{\Psi_0}(\lam)  &=& \limth \frac{i}{N} \langle {\Psi_0} |
T^{-1} \! \left( \lam + \tfrac{i\eta}{2}\right) \partial_\lam T \!
\left( \lam +\tfrac{i\eta}{2} \right) | {\Psi_0} \rangle,\nn
&=&\sum_{k=0}^\infty \frac{\lambda^{k}}{k!}h^{(k+1)}.
\eea

Given the GGE density matrix, a ``microcanonical'' description of the steady
state can be obtained by performing a generalized Thermodynamic Bethe
Ansatz (GTBA) \cite{2012_Mossel_JPA_45,2012_Caux_PRL_109}, see SM for
a brief summary. This results
in a representative eigenstate $|\boldsymbol{\rho}^{\Psi_0}_\text{GGE}
\rangle$ labeled by root density functions
$\boldsymbol{\rho}^{\Psi_0}_\text{GGE} $, which has the property that
for any local operator $\cO$
\begin{equation}
\text{Tr}(\cO \hat{\varrho}_\text{GGE}) = \big\langle\brho^{\Psi_0}_\text{GGE} \big| \cO \big| \brho^{\Psi_0}_\text{GGE}\big\rangle.
\end{equation}
Within the GTBA formalism macrostates can be described either by
root densities of particles, or by densities of holes. Holes can be, loosely speaking,
understood as analogues of unoccupied states in models of free fermions. In terms of the
latter the state $|\boldsymbol{\rho}^{\Psi_0}_\text{GGE}\rangle$ is
parametrized in terms of the set of positive functions
$\{\rho^{\Psi_0}_{n,h}\}$. In
\cite{2014_Wouters_PRL_113,2014_Brockmann_JSTAT_P12009} it was found
that the initial data \fr{GGE_fixbeta} directly determines the hole density
of 1-strings (i.e. vacancies of unbound states), according to the remarkable identity 
\begin{equation}
\rho_{1,h}^{\Psi_0} (\lambda) = a_1 (\lambda) + \frac{1}{2\pi} \left[ \Omega^{\Psi_0} (\lambda + \tfrac{i\eta}{2}) + \Omega^{\Psi_0} (\lambda - \tfrac{i\eta}{2}) \right].
\label{eq:rho1h}
\end{equation}
All other hole densities are fixed by the maximum entropy principle
under the constraints \fr{GGE_fixbeta}.

\paragraph*{Quench Action treatment.}
The above GGE treatment should be compared to an independent calculation
using the QA method \cite{2013_Caux_PRL_110}. For a generic
quench problem, given an initial state $|\Psi_0 \rangle$, the
time-dependent expectation value of a generic local observable $\cO$
can be expressed as a double Hilbert space summation
\begin{equation} \label{eq:expect1}
\left\langle \Psi (t) \right| \cO \left| \Psi (t) \right\rangle = 
\sum_{\blam,\blam'} e^{-S_{\blam}^*-S_{\blam'}}e^{i(\omega_{\blam} - 
\omega_{\blam'})t} \langle \blam| \cO |\blam'\rangle,
\end{equation}
where $S_{\blam} = - \ln \left\langle \blam | \Psi_{0}
\right\rangle$. Here, $|\blam\rangle$ are eigenstates of the
Hamiltonian driving the post-quench time evolution. Exploiting the
fact that in the themodynamic limit, the summation over eigenstates
can be written as a functional integral over root densities, which can
be evaluated in a saddle-point approximation (becoming exact in
the thermodynamic limit), one finds in particular that the
steady-state expectation values of observables a long time after the
quench can be obtained as 
\begin{equation}
\lim_{t \to \infty}\limth \left\langle \Psi (t) \right| \cO \left| \Psi (t) \right\rangle  = 
\big\langle\brho^{\Psi_0}_{\text{QA}} \big| \cO \big| \brho^{\Psi_0}_\text{QA}\big\rangle.
\label{eq:spev}
\end{equation}
Here $\big|\brho^{\Psi_0}_\text{QA}\big\rangle$ is an eigenstate
minimizing the QA $S_{\text{QA}}[\boldsymbol{\rho}] = 2
S[\boldsymbol{\rho}] - S_{\text{YY}}[\boldsymbol{\rho}]$, where
$S[\boldsymbol{\rho}] = \limth \text{Re}\, S_\blam$ is the extensive
real part of the overlap coefficient in the thermodynamic limit and
$S_{\text{YY}}[\boldsymbol{\rho}]$ is the Yang-Yang entropy  
of the state \cite{GaudinBOOK,KorepinBOOK,TakahashiBOOK}.
For the N{\'e}el to $XXZ$ quench, the exact overlaps were obtained in
\cite{2014_Brockmann_JPA_47a} and used in
\cite{2014_Wouters_PRL_113,2014_Brockmann_JSTAT_P12009} to obtain the
exact saddle-point densities $\boldsymbol{\rho}^{\Psi_0}_{QA}$
representing the steady state. Crucially, one finds 
\cite{2014_Wouters_PRL_113,2014_Pozsgay_PRL_113} that
$\boldsymbol{\rho}^{\Psi_0}_{\text{GGE}}
\neq\boldsymbol{\rho}^{\Psi_0}_{\text{QA}}$, which in turn leads to 
different predictions for physical properties such as spin-spin
correlators. This demonstrated that the ultra-local GGE does not
correctly describe the steady state after a generic quantum quench in the XXZ
chain. 

\paragraph*{Constructing a ``quasi-local'' GGE.}
Very recently \cite{Ilievski_arxiv150605049} 
(see also 
\cite{2011_Prosen_PRL_106,2013_Prosen_PRL_111,2013_Ilievski_CMP_318,2014_Prosen_NPB_886,2014_Pereira_JSTAT_P09037,2015_Mierzejewski_PRL_114})
hitherto unknown conserved charges of the isotropic ($\Delta = 1$)
Heisenberg model were discovered. These operators are not local in the sense that they 
cannot be represented as a spatially homogeneous sum of finitely
supported densities, but rather quasi-local, meaning
\cite{Ilievski_arxiv150605049} that their Hilbert--Schmidt norms scale
linearly with system size and their overlaps with locally-supported
operators become independent of $N$ in the limit of large system
size. Moreover, they are linearly independent from the known local
charges generated from the spin-$1/2$ transfer matrix. Until now, the
impact of these charges on local physical observables has not been quantified. 

Our first step is to construct a family of quasi-local conserved
charges for $\Delta\geq 1$ by generalizing the procedure of
\cite{Ilievski_arxiv150605049}. The starting point is the $q$-deformed
L-operator, 
\begin{align}
& \quad L(z,s) =\frac{1}{\sinh \eta }
\Big(\sinh (z) \cosh{(\eta s^{z})}\otimes \sigma^{0} \\
&+ \! \cosh (z) \sinh{(\eta s^{z})}\otimes \sigma^{z} \!\! +\sinh (\eta) (s^{-}\otimes \sigma^{+} \!\! +s^{+}\otimes \sigma^{-})\Big), \notag
\end{align}
whose auxiliary-space components are given by $q$-deformed spin-s representations with $s=\tfrac{1}{2},1,\tfrac{3}{2},\ldots$, obeying commutation relations
$[s^{+},s^{-}]=[2s^{z}]_{q}$, $[s^{z},s^{\pm}]=\pm s^{\pm}$
and acting in a $(2s+1)$-dimensional irreducible representation
$\mathcal{V}_{s}\cong \CC^{2s+1}={\rm lsp}\{\ket{k};k=-s,\ldots,s\}$, 
\begin{equation}
s^{z}\ket{k}=k\ket{k},\quad s^{\pm}\ket{k}=\sqrt{[s+1\pm k]_{q}[s\mp k]_{q}}\ket{k\pm 1},
\end{equation}
with $[x]_q =\sinh{(\eta x)}/\sinh(\eta)$. 
By means of higher-spin auxiliary (fused) transfer matrices defined via ordered products of L-operators
\begin{equation}
T_{s}(z)={\rm Tr}_{a}\left[  L_{a,1}(z,s) \ldots L_{a,N}(z,s)\right] ,
\label{Tsz}
\end{equation}
(where $T_{1/2}(z) \equiv T(z)$ was used in (\ref{ultralocal}))
we define families of conserved operators
\begin{equation}
X_{s}(\lambda) =\tau^{-1}_{s}(\lambda)\{T_{s}(z^{-}_{\lambda})T^{\prime}_{s}(z^{+}_{\lambda})\},
\quad z^{\pm}_{\lambda}=\pm \frac{\eta}{2}+i \lambda,
\label{eq:X_def}
\end{equation}
with $T^{\prime}_{s}(z)\equiv \partial_{z}T_{s}(z)$ and $\{\bullet\}$ denoting the traceless part.
The normalization reads
$\tau_{s}(\lambda)=f(-(s+\tfrac{1}{2})\eta+i \lambda)f((s+\tfrac{1}{2})\eta+i \lambda)$ with $f(z)=(\sinh{(z)}/\sinh{(\eta)})^{N}$. In 
\cite{Ilievski_arxiv150605049} it was shown for the isotropic case that these charges are quasi-local for all $s=\tfrac{1}{2},1,
\tfrac{3}{2},\ldots$ and $\lambda \in \mathbb{R}$. A rigorous proof for general $\Delta>1$ is currently under
construction~\footnote{M. Medenjak, E. Ilievski and T. Prosen, to be published}.

A central piece of our work is the extraction of the thermodynamically
leading part of the quasi-local charges $\{X_{s}\}_{s=1/2}^{\infty}$ when operating on an arbitrary
Bethe state. It proves useful to resort to the so-called  
fusion relations \cite{2010_Bazhanov_JSTAT_P11002, 2007_Bazhanov_NPB_775,1999_Suzuki_JPA_32,1992_Kluemper_PA_183} (T-system) 
for higher-spin transfer matrices,
\begin{align}\label{T-system}
T_{s}(z+\tfrac{\eta}{2})T_{s}(z-\tfrac{\eta}{2})
 & =f(z+(s+\tfrac{1}{2})\eta)f(z-(s+\tfrac{1}{2})\eta)\nonumber \\
& +T_{s-1/2}(z)T_{s+1/2}(z),
\end{align}
with the initial condition $T_{0}(z)\equiv f(z)$.
There exists a closed-form solution to the above recurrence relation
in terms of Baxter's $Q$-operators \cite{2007_Bazhanov_NPB_775}
\begin{align} \label{eq:closed_form}
T_{s}(z)&=Q(z +  (s+\tfrac{1}{2})  \eta )Q(z-(s+\tfrac{1}{2})  \eta) \\
&\times \sum_{\ell=0}^{2s}\frac{f(z+(\ell -s)  \eta)}{Q(z+(\ell -s-\tfrac{1}{2})  \eta)Q(z+(\ell -s+\tfrac{1}{2})  \eta)}. \notag
\end{align}
The eigenvalues of the $Q$-operators (in what follows, in view of commutations $\left[ T_s (z_1), Q(z_2) \right] = 0 ~\forall s, z_i 
\in \CC$, we slightly abuse notation by using the same symbol for an operator and its eigenvalue) are determined by the position of 
Bethe roots $Q(z) =\prod_{k=1}^{M}\sinh{(z+i\lambda_k)}$. A key observation is
that, in the thermodynamic limit, the spin-$s$ transfer matrix
evaluated at $z_\lambda^-$ ($z_\lambda^+$) is simply given by the
$\ell =0$ ($\ell =2s$) term in the sum in
Eq.~\eqref{eq:closed_form}. This then gives
\begin{align} \label{eq:limth_transfer}
\limth T_{s}(z^{\pm}_{\lambda})=\limth f(\pm (s+\tfrac{1}{2})\eta+ i \lambda)
\frac{Q(\mp s\eta+ i \lambda)}{Q(\pm s\eta+ i \lambda)}.\nonumber \\
\end{align}
The latter analysis is consistent with
$\limth \tau^{-1}_{s}(\lambda)T_{s}(z^{-}_{\lambda})T_{s}(z^{+}_{\lambda})=\one$,
representing a thermodynamic version of an inversion identity
(see \cite{Ilievski_arxiv150605049}) that can be proven in an entirely
operatorial way, without making reference to the Bethe eigenstates. 
At this point it is convenient to define modified conserved operators
\begin{equation}
\widehat{X}_{s}(\lambda):=T^{(-)}_{s}(z^{-}_{\lambda})T^{(+)\prime}_{s}(z^{+}_{\lambda}),
\end{equation}
where $T^{(\pm)}_{s}(z)$ have the same structure as \fr{Tsz} but
involve L-operators $L^{(\pm)}(z,s) =L(z,s) \sinh{(\eta)}/[\sinh{(z\pm s\eta)}]$.
In thermodynamic limit a quasi-local conserved operator $\widehat{X}_{s}(\lambda)$ only differs
from $X_{s}(\lambda)$ by a multiple of identity,
$\widehat{X}_{s}(\lambda)=X_{s}(\lambda)+t_{s}(\lambda)\one$, with
$t_{s}(\lambda)=\tfrac{2s}{2s+1}\tfrac{\sinh{((2s+1)\eta)}}{\sinh^{2}{(\eta)}}\tau^{-1}_{s}(\lambda)$.
We can now define a two-parameter family of conserved charges 
\bea
H^{(n+1)}_s=\frac{1}{n!}\partial^n_\lambda \widehat{X}_s(\lambda) \Big|_{\lambda=0}.
\eea
By construction we have $[H^{(n)}_s,H^{(m)}_{s'}]=0$ and
$\{H^{(n)}_{1/2}\}_{n=1}^{\infty}$ precisely recover the ultra-local conservation
laws \fr{ultralocal}. We are thus in a position to define the density
matrix of our GGE. It is given by
\be
\hat{\varrho}_{\rm GGE}=\frac{1}{Z}\exp\left[
-\sum_{n,s=1}^\infty\beta_n^sH^{(n)}_{s/2}\right],
\label{ourGGE}
\ee
where the Lagrange multipliers $\beta_n^s$ are fixed by initial
conditions of the form \fr{GGE_fixbeta}. Our assertion is that 
\fr{ourGGE} provides a correct description of the stationary behaviour
after a general quench to the spin-1/2 XXZ chain (in the regime $\Delta
\geq 1$). In order to prove this it suffices to establish that the
initial values of our conserved charges uniquely specify a macrostate.

Let us now derive the main result of our
Letter. Analogously to what was found in \cite{2014_Wouters_PRL_113,
  2014_Brockmann_JSTAT_P12009} for the ultra-local charges, the values of
the quasi-local charges associated with a spin-$s$ transfer matrix are
in 1-to-1 correspondence with functions $\rho^{\Psi_{0}}_{2s,h}(\lambda)$,
which in turn specify (see SM) a unique macrostate (namely the GGE saddle-point state).

Our starting point is the following expression for the spectrum of
$\{\widehat{X}_s\}_{s=1/2}^{\infty}$, valid for large system size
(cf. Eq.~\eqref{eq:limth_transfer})
\begin{align}\label{eq:spectrum}
 \widehat{X}_{s}(\lambda)  &=
-i \partial_{\lambda}\log\frac{Q(-s\eta+i \lambda)}{Q(s\eta+i \lambda)}+o(N) \\
&=\sum_{k=1}^{M}\frac{2\sinh{(2s\eta)}}{\cos(2(\lambda_{k}+ \lambda))-\cosh(2s\eta)}+o(N), \notag
\end{align}
Starting from Eq.~(\ref{eq:spectrum}), working in the thermodynamic limit under the string hypothesis and making use of Bethe equations, one arrives at (see SM)
\begin{equation}
\rho_{2s,h}^{\Psi_0} (\lambda) = a_{2s} (\lambda) + \frac{1}{2\pi} \left[ \Omega^{\Psi_0}_{s} (\lambda + \tfrac{i\eta}{2}) + \Omega^{\Psi_0}_{s} (\lambda - \tfrac{i\eta}{2}) \right], 
\label{eq:rho2sh}
\end{equation}
where $s=\tfrac{1}{2},1,\tfrac{3}{2},\ldots$. The right-hand side 
of \fr{eq:rho2sh} is determined by the expectation values of the
quasi-local charges on the initial state,  
\begin{equation}
\Omega^{\Psi_0}_{s} (\lambda) = \limth \frac{\langle \Psi_0 | \widehat{X}_s (\lambda) | \Psi_0 \rangle}{N}.
\label{eq:Omegas}
\end{equation}
This is a generalization of Eq.~\eqref{eq:rho1h} to arbitrary
spin. Note the remarkable fact that this correspondence is valid for a
generic initial state $|\Psi_0\rangle$. As a consequence, the family
of quasi-local charges $\{\widehat{X}_s \}_{s=1/2}^\infty$
completely determines the postquench stationary state through the GGE
and gives the latter's predictions identical to those coming from the
QA.

\paragraph*{N\'eel quench.}
An explicit example of our construction if provided by the quench from
the N{\'e}el state 
\begin{equation}\label{eq:Neel_state}
	|\Psi_0\rangle = \frac{1}{\sqrt{2}}\left(\left|\uparrow\downarrow \uparrow\downarrow\ldots\right\rangle+
	\left|\downarrow\uparrow\downarrow\uparrow \ldots\right\rangle\right).
\end{equation}
Here the root distributions characterizing the stationary state have
been previously determined by a QA calculation
\cite{2014_Wouters_PRL_113,2014_Brockmann_JSTAT_P12009}. In order to
demonstrate that our GGE recovers these known results we need to
compute the generating functions \eqref{eq:Omegas}. Here we can repeat
the logical steps from the calculation for $s=1/2$ in
\cite{2013_Fagotti_JSTAT_P07012,2014_Fagotti_PRB_89} by studying the spectrum of 
associated auxiliary transfer matrices. This calculation can be found in the SM.
Substituting the results obtained in this way into \eqref{eq:rho2sh}
gives perfect agreement with the known QA results.

\begin{figure}
\includegraphics[width=\columnwidth]{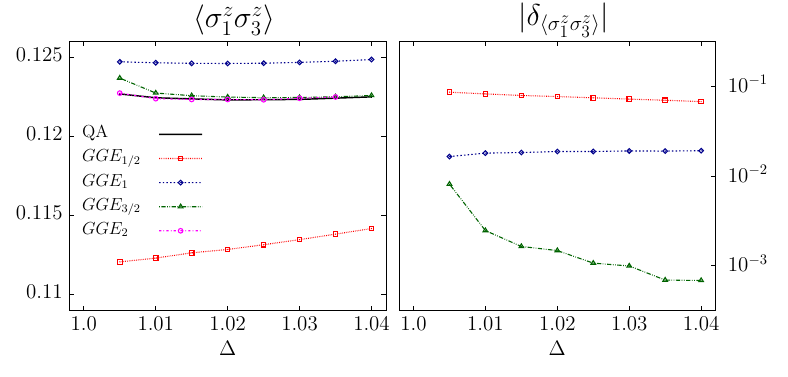}
\caption{Comparison of methods: QA method versus improved GGE predictions. Colored lines pertain to the refined GGE calculation with systematic addition of higher-spin families of quasi-local charges $\{ H_s^{(n)} \}$ for the local correlation function
$\langle \sigma^{z}_{1} \sigma^{z}_{3}\rangle$ in the regime $\Delta \gtrsim 1$ (left panel).
Labels in $GGE_{\bar{s}}$ indicate the maximal auxiliary spin $\bar{s}$ for the charges $\{ H_{s}^{(n)} \}$ being included in the GGE computation.
The right panel displays the relative differences
$\delta_{\langle \sigma^z_1 \sigma^z_3 \rangle} = (\langle \sigma^{z}_{1}\sigma^{z}_{3} \rangle_{QA} - \langle \sigma^{z}_{1}\sigma^{z}_{3} \rangle_{GGE_{\bar{s}}})/\langle \sigma^{z}_{1}\sigma^{z}_{3} \rangle_{QA}$ in logarithmic scale.
We used the mapping between correlation functions and the set of densities $\boldsymbol{\rho}$ given
in \cite{Mestyan_JSTAT_2014}.}
\label{fig:GGEtoQA}
\end{figure}

\paragraph*{Towards a truncated GGE.} In \cite{2013_Fagotti_PRB_87} it
was argued that for the purpose of describing finite subsystems in the
thermodynamic limit ultra-local GGEs can be truncated by retaining only
a finite number of the ``most local'' conserved charges. An obvious
question is whether a similar logic can be applied to our quasi-local
GGE. As a first step towards understanding this issue, we have
calculated the next-nearest spin correlation function in the steady state
after a  N{\'e}el-to-XXZ quench for several GGEs truncated in the $s$
direction. In Fig. \ref{fig:GGEtoQA} we show the results of these
calculations for $\Delta\gtrsim 1$. The data clearly shows that adding
subsequent families of quasi-local charges results in a rapid
convergence of the corresponding truncated GGE result to the exact
value.

\paragraph*{Conclusions.}
We have shown how to construct an exact GGE describing the stationary
state after generic quantum quenches to the spin-1/2 Heisenberg XXZ
chain. Our GGE is built from an extended set of local and
quasi-local charges. We have shown that our construction resolves
previously observed discrepancies between predictions for steady state
expectation values by an exact QA treatment on the one hand,
and a GGE restricted to ultra-local charges obtained from the transfer
matrix of the spin-1/2 chain on the other hand. Our results provide
unambiguous proof that the recently discovered quasi-local charges have
a non-negligible impact on the relaxation processes of
strongly-interacting many-body quantum systems in one dimension.

\paragraph*{Acknowledgements.}
We warmly acknowledge discussions with M. Brockmann and thank him for careful reading of the manuscript.
BW, JDN and JSC acknowledge support from the Foundation for
Fundamental Research on Matter (FOM) and the Netherlands Organisation
for Scientific Research (NWO). FHLE is supported by the EPSRC under
grants EP/J014885/1 and EP/I032487/1. This work forms part of the
activities of the Delta Institute for Theoretical Physics (D-ITP).
TP acknowledges support by grants P1-0044, N1-0025 and J1-5439 of
Slovenian Research Agency.

\bibliography{Neel_XXZ_GGE_fix}

\setcounter{equation}{0}
\makeatletter
\renewcommand{\theequation}{S\arabic{equation}}

\newpage

\begin{widetext}

\pagebreak

\begin{center}
\textbf{\large Supplementary Material:\\ Complete Generalized Gibbs Ensemble in an Interacting Theory}
\end{center}

\end{widetext}

\twocolumngrid

We begin by recalling some of the fundamental equations of the Bethe
Ansatz solution of the $XXZ$ chain, together with few results obtained
in \cite{2014_Wouters_PRL_113,2014_Brockmann_JSTAT_P12009} which are
used in the bulk of our paper.  
\subsection{Thermodynamic limit of Bethe equations}
In the thermodynamic limit, the Bethe equations for the $XXZ$ chain read \cite{GaudinBOOK,KorepinBOOK,TakahashiBOOK}
\begin{equation}
	\rho_{n,t}(\lam) \ =\ a_{n}(\lam) - \sum_{m=1}^{\infty}  (a_{nm} \ast \rho_{m}) (\lam),
\end{equation}
for $n\geq1$, where $\rho_{n,t}(\lambda) = \rho_{n}(\lambda) + \rho_{n,h}(\lambda)$ and $\rho_n, \rho_{n,h}$ are respectively the particle and hole densities of $n$-strings. 
The convolution is defined by
$(f\ast g)\, (\lam) = \int_{-\pi/2}^{\pi/2} \mathrm{d}\mu \, f(\lam-\mu)\, g(\mu)$.
The kernels are
\begin{align} 
\label{eq:kernelXXZsum}
	a_{nm}(\lam) &=  (1-\delta_{nm}) a_{|n-m|}(\lam) + 2 a_{|n-m| + 2}(\lam) + \ldots \nonumber \\
& + 2 a_{n+m-2}(\lam) + a_{n+m}(\lam),
\end{align}
with $a_n$ defined in (\ref{eq:an}) in the main part of the manuscript.
A convenient rewriting is in the decoupled form \cite{TakahashiBOOK}
\begin{equation} \label{eq:BTGthlim_fact_a}
	\rho_{n}(1 + \eta_{n}) = s \ast (\eta_{n-1}\rho_{n-1} + \eta_{n+1}\rho_{n+1}),
\end{equation}
for $n\geq 1$, where $\eta_{n} \equiv \rho_{n,h}/\rho_{n}$. The $\lam$-dependence is left implicit and we use the conventions $\eta_0(\lam)=1$ and $\rho_{0}(\lam)=\delta(\lam)$. The kernel in \eqref{eq:BTGthlim_fact_a} reads 
\begin{equation} \label{eq:defs}
	s(\lam) = \frac{1}{2\pi} \sum_{k\in\mathbb{Z}} \frac{e^{-2i k\lam}}{\cosh(k\eta)} \epp
\end{equation}
The set $\brho=\{\rho_n\}_{n=1}^\infty$ represents an ensemble of states with Yang-Yang entropy
\begin{align} \label{eq:YYentropyXXZ}
	S_{YY} \left[ \brho \right] = N \sum_{n=1}^{\infty}  \int_{-\pi/2}^{\pi/2} \mathrm{d}\lam \left[ \rho_{n,t}(\lambda) \ln\rho_{n,t}(\lambda) \right.  \nonumber \\ \left. 
- \rho_{n}(\lambda) \ln \rho_{n}(\lambda) - \rho_{n,h}(\lambda)\ln\rho_{n,h}(\lambda) \right]\epp
\end{align}
An important point to bear in mind is that the Bethe equations (\ref{eq:BTGthlim_fact_a}) relate the set of densities $\brho$ to the set of hole densities $\brho_h$. Knowing one of these two sets is thus sufficient to completely determine a given state. This point is crucial to understand the effects of constraints coming from ultra-local and quasi-local charges, as is explained below.

\subsection{GTBA for the GGE}
The generalized TBA for the GGE based on local charges proceeds as a standard TBA, but now with the effect of additional charges beyond the Hamiltonian being taken into account by additional parameters $\beta_n$ (chemical potentials) in the GGE density matrix. By applying the standard maximal entropy reasoning using these constraints results in the GTBA equations \cite{2014_Wouters_PRL_113,2014_Brockmann_JSTAT_P12009}
\begin{equation}\label{eq:GTBAGGE_a}
	\ln(\eta_n) = -\delta_{n,1} (s \ast d) + s\ast\left[\ln(1+\eta_{n-1}) + \ln (1+\eta_{n+1}) \right] \epc
\end{equation}
for $n\geq 1$, where $\eta_0(\lam)=0$ and $s(\lambda)$ is defined in \eqref{eq:defs}.
The driving term originating from ultra-local charges is remarkably only present in the first integral equation and is specified by the chemical poten\-tials~$\beta_m$,~$m \geq 2$,
\begin{equation}\label{eq:GTBAGGE_b}
	d(\lam) = \sum_{k \in \mathbb{Z}} e^{-2ik\lam} \sum_{m=2}^\infty \beta_m \sinh^{m-1}(\eta) (ik)^{m-2} \epp
\end{equation}
As shown in \cite{2014_Wouters_PRL_113,2014_Brockmann_JSTAT_P12009}, the ultra-local charges associated to the spin-$1/2$ transfer matrix completely fix the density of holes of 1-strings to $\rho^{\Psi_0}_{1,h}$, but leave all higher hole density functions $\brho_{n,h}$, with $n \geq 2$, undetermined. As explained in detail in \cite{2014_Brockmann_JSTAT_P12009}, the GTBA system of (\ref{eq:GTBAGGE_a}) and (\ref{eq:BTGthlim_fact_a}) for the GGE can then be solved (using the constraint $\rho_{1,h} = \rho^{\Psi_0}_{1,h}$ to eliminate the unknown driving term in (\ref{eq:GTBAGGE_a})).

\subsection{GTBA for the Quench Action}
In the case of the QA treatment, the GTBA equations take the form
\begin{equation}\label{eq:TBA_XXZ_fact_equation}
	\ln(\eta_n) = d_n + s \ast \big[\ln(1+\eta_{n-1})+\ln(1+\eta_{n+1})\big]\epc
\end{equation}
where $n\geq 1$. The driving terms are given by the exact overlaps of Bethe states with the initial state. In the specific case of the N{\'e}el quench, these are given by 
\begin{align}\label{eq:TBA_XXZ_fact_driving}
	d_n(\lam) &= \sum_{k\in\mathbb{Z}} e^{-2ik\lam}\frac{\tanh(k\eta)}{k}\left[(-1)^n-(-1)^k\right] 
\nonumber \\ &= (-1)^n\ln\left[\frac{\vartheta_4^2(\lambda)}{\vartheta_1^2(\lambda)}\right] +\ln\left[\frac{\vartheta_2^2(\lambda)}{\vartheta_3^2(\lambda)}\right]\epc
\end{align}
where $\vartheta_j$, $j=1,\ldots,4$, are Jacobi's $\vartheta$-functions with nome $e^{-2\eta}$.

Note the difference between this GTBA and the one associated to the ultra-local GGE: from the exact QA treatment one obtains GTBA equations with driving terms at all string lengths $n$, as a result yielding a different set of the steady state densities.

\subsection{Relating quasi-local charges and hole densities}
In the main text in \eqref{eq:spectrum} we provided the expectation values of the quasi-local charges on
Bethe eigenstates in the limit of large system size. Strictly in the thermodynamic limit one obtains
\begin{align}
& \quad \,\limth \frac{1}{N} \bra{\blam}\widehat{X}_{s}(\mu)\ket{\blam} \notag \\
&= \limth \frac{1}{N} \sum_{k=1}^{M}\frac{2\sinh{(2s\eta)}}{\cosh{(2(z_k-i \mu))}-\cosh{(2s\eta)}} \\
&= -2 \pi \sum_{n=1}^\infty \int_{-\pi/2}^{\pi/2} d \lambda \, \rho_n(\lambda) \sum_{j=1}^{\text{min}(n,2s)} a_{|n-2s|-1+2j} (\lambda +\mu) , \label{eq:spectrum_strict_limth}
\end{align}
with $a_n$ defined in ~\eqref{eq:an}. In the last equality we accounted for $z_k = -i \lam_k$, used the string hypothesis and the fact that the expectation values can be written as
\begin{align}
& \quad \,  \bra{\blam}\widehat{X}_{s}(\mu)\ket{\blam} \notag \\
& = -i\,\partial_\mu \ln \left( \prod_{k=1}^M \frac{\sin(\lam_k + \mu + i s \eta)}{\sin(\lam_k + \mu - i s \eta)} \right) + o (N).
\end{align}
Using conventions for the Fourier transform
\begin{align}
	\hat{f}(k) &= \int_{-\pi/2}^{\pi/2}\mathrm{d}\lambda \, e^{2ik\lam}f(\lam)\epc\quad k\in\mathbb{Z} \epc \\
	f(\lam) &=  \frac{1}{\pi}\sum_{k\in\mathbb{Z}} e^{-2ik\lam}\hat{f}(k)\epc\quad \lambda\in[-\tfrac{\pi}{2}, \tfrac{\pi}{2}) \epc \label{eq:inverse_FourierTransform}
\end{align}
one can map Eq.~\eqref{eq:spectrum_strict_limth} to Fourier space,
\begin{align}
& \quad \,\limth \frac{1}{N} \bra{\blam}\widehat{X}_{s}(\mu)\ket{\blam}\\
&= -2 \sum_{k\in\mathbb{Z}} e^{-i2k\mu}  \sum_{n=1}^\infty  \hat{\rho}_n(k) \sum_{j=1}^{\text{min}(n,2s)} e^{-|k|\eta (|n-2s| -1 +2j)} , \notag
\end{align}
using that $\hat{a}_n(k) = e^{-|k|\eta n}$. By performing the sum over $j$ and using that $|n-2s| + 2\,\text{min}(n,2s) = n+2s$, one finds
\begin{align}
& \quad \,\limth \frac{1}{N} \bra{\blam}\widehat{X}_{s}(\mu)\ket{\blam}\\
&=  \sum_{k\in\mathbb{Z}}  \frac{e^{-i2k\mu}}{\sinh(|k|\eta)}  \sum_{n=1}^\infty  \hat{\rho}_n(k) \left( e^{-|k|\eta (n+2s)} - e^{-|k|\eta |n-2s|} \right) . \notag
\end{align}
Using the thermodynamic Bethe equations (cf.~\eqref{eq:BTGthlim_fact_a}) in Fourier space,
\begin{equation}
\hat{\rho}_{n,t} (k) = \frac{1}{2\cosh(k\eta)} \left( \hat{\rho}_{n-1,h}(k) +\hat{\rho}_{n+1,h}(k) \right) ,
\end{equation}
where $\hat{\rho}_{0,h}(k)=1$, one can observe a cancellation of all terms with an exception of an expression given solely
in terms $\rho_{2s,h}$:
\begin{align} \label{eq:in_terms_of_rho2s}
& \quad \,\limth \frac{1}{N} \bra{\blam}\widehat{X}_{s}(\mu)\ket{\blam} \notag \\
&=  \sum_{k\in\mathbb{Z}} \frac{e^{-i2k\mu}}{\cosh(k\eta)} \left( \hat{\rho}_{2s,h}(k) - e^{-2s|k|\eta} \right)  . 
\end{align}
The quasi-local conservation laws make the left-hand side of Eq.~\eqref{eq:in_terms_of_rho2s} equal to the generating function of the charges on the initial state as stated previously in \eqref{eq:Omegas}, leading to
\begin{equation}
\hat{\Omega}^{\Psi_0}_{s} (k) \frac{\cosh(k\eta)}{\pi}  =  \hat{\rho}_{2s,h}(k) - e^{-2s|k|\eta} .
\end{equation}
Taking the inverse Fourier transform produces the main result of our Letter, namely the identification given by \eqref{eq:rho2sh}.

\subsection{Truncated GGE}
For practical reasons it is useful to determine a GGE ensemble by including only a finite number $\bar{s}$ of quasi-local charges $\{ \widehat{X}_s\}_{s=1}^{\bar{s}}$. Using Eq. \eqref{eq:rho2sh} for $s=1, \ldots , \bar{s}$ we fix the distributions of holes $\rho_{n,h}= \rho_{n,h}^{\Psi_0}$ for string of lengths $n = 1, \ldots , 2\bar{s}$. These restrictions can be in turn used as a driving term for the following GTBA equations (analogously to what has been done for the case $\bar{s} = 1/2$ in \cite{2014_Wouters_PRL_113,2014_Brockmann_JSTAT_P12009,2014_Pozsgay_JSTAT_P09026})
\begin{equation}
	\ln(\eta_n) =  s \ast \big[\ln(1+\eta_{n-1})+\ln(1+\eta_{n+1})\big] \qquad  n> 2\bar{s} \epc
\end{equation}
 with 
\begin{equation}
\eta_{n} = \frac{\rho_{n,h}^{\Psi_0} }{\rho_n}  \qquad  n \leq 2 \bar{s} \epc
\end{equation}
where the functions $\rho_n$ solve the Bethe equations \eqref{eq:BTGthlim_fact_a}. For any $\bar{s} \geq 1/2$ we then
apply an iterative procedure to find a solution for all the $\{ \eta^{\bar{s}}_n\}_{n=1}^{\infty}$ and all the $\{ \rho^{\bar{s}}_n\}_{n=1}^{\infty}$, ultimately leading to the results shown in figure \ref{fig:GGEtoQA}.

\subsection{N{\'e}el initial state}
As the N\'eel state is a
simple product state we can evaluate all scalar products in the
``quantum'' spaces pertaining to the physical spin-1/2 degrees of
freedom. This leaves us with a staggered product of diagonal components of an
auxiliary two-channel L-matrix $\mathbb{L}_{s}(z_{1},z_{2})=L^{(-)}_{a_{1}}(z_{1},s)L^{(+)}_{a_{2}}(z_{2},s)$, producing a transfer matrix
$\mathbb{T}_{s}(z_{1},z_{2})$ operating on two copies of auxiliary spin-$s$ spaces
$ \mathcal{V}_{s}\otimes \mathcal{V}_{s}$,
\begin{equation}
\mathbb{T}_{s}(z_{1},z_{2})=\mathbb{L}^{\ua \ua}_{s}(z_{1},z_{2})\mathbb{L}^{\da \da}_{s}(z_{1},z_{2}).
\end{equation}
This is a local unit of the \squote{boundary partition function}
$Z_{s}(z_{1},z_{2})=\limth N^{-1}{\rm Tr}_{a}\mathbb{T}_{s}(z_{1},z_{2})^{N/2}$, whose large-$N$ limit on
$\mathcal{D}:=\{(z^{-}_{\lambda},z^{+}_{\lambda});\lambda \in \RaR\}\subset \CC^{2}$ is dominated by a non-degenerate unit eigenvalue
$\Lambda_{s}(z^{-}_{\lambda},z^{+}_{\lambda})=1$ of $\mathbb{T}_{s}(z_{1},z_{2})|_{\mathcal{D}}$, implying
\begin{equation}
\Omega^{\text{N\'eel}}_{s}(\lambda)=\partial_{z_{2}}Z_{s}(z_{1},z_{2})|_{\mathcal{D}}
=\frac{1}{2}[\partial_{z_{2}}\Lambda_{s}(z_{1},z_{2})]_{\mathcal{D}}.
\end{equation}
It also helps noticing that $\mathbb{T}_{s}(z_{1},z_{2})$ enjoys a
$U(1)$-symmetry, with the leading eigenvalue always residing in the
largest $(2s+1)$-dimensional subspace. Closed-form results can be
readily obtained in the cases $s=\tfrac{1}{2},1$
\begin{align}
\Omega^{\text{N\'eel}}_{1/2}(\lambda)&=\frac{-\sinh{(2\eta)}}{1-2\cos{(2\lambda)}+\cosh{(2\eta)}},\\
\Omega^{\text{N\'eel}}_{1}(\lambda)&=\frac{2\sinh{(3\eta)}}{3\cos{(2\lambda)}-\cosh{(\eta)}-2\cosh{(3\eta)}}.
\end{align}
suppressing $\Omega^{\text{N\'eel}}_{s}(\lambda)$ for higher spins $s \geq 3/2$ which become quickly cumbersome expressions. For 
practical implementation of the truncated GGE it however suffices to evaluate them numerically. For a class of initial states which 
are given in the Matrix Product State form this can be done efficiently by e.g. employing the method outlined in 
\cite{2014_Fagotti_PRB_89}.

\end{document}